\documentclass[aps,pra,preprint,groupedaddress,superscriptaddress]{revtex4-2}

\usepackage{graphicx}
\usepackage{dcolumn}
\usepackage{bm}
\usepackage{amsmath}
\usepackage{amssymb}
\usepackage{braket}
\usepackage[hidelinks]{hyperref}

\begin{document}
	
	\title{Photon-Conditioned Squeezed States for Directional Displacement Response in Continuous-Variable Photonics}
	
	\author{Boris Kiefer}
	\email{bkiefer@nmsu.edu}
	\affiliation{Department of Physics, New Mexico State University, Las Cruces, NM, USA}
	
	\author{Olivier Pfister}
	\affiliation{Department of Physics, The University of Virginia, Charlottesville, VA, USA}
	\affiliation{Charles L. Brown Department of Electrical and Computer Engineering, University of Virginia, 351 McCormick Road, Charlottesville, VA 22903, USA}
	
	\date{\today}
	
\begin{abstract}
	Squeezed Fock states, photon-subtracted squeezed states, and optical cat
	states are established non-Gaussian resources in continuous-variable quantum
	optics. Here we compare these known state families from a task-oriented
	perspective: matched mean photon number, scalar Wigner negativity, and
	directional displacement-fidelity response. Starting from squeezed vacuum,
	single-photon subtraction prepares a state proportional to
	\(S(r,\theta)\ket{1}\), while two-photon subtraction prepares an even-parity
	squeezed Fock superposition rather than a pure \(S(r,\theta)\ket{2}\).
	We benchmark photon-conditioned squeezed states against Fock and coherent-cat
	references using the integrated Wigner negativity \(\delta\), the
	energy-normalized metric \(\delta/\langle n\rangle\), and fidelity-threshold
	displacement radii \(R_F(\phi)\). Cat benchmarks remain strong scalar
	Wigner-negativity resources, whereas photon-conditioned squeezed states
	provide an origin-centered alternative with tunable anisotropic displacement
	response. In particular, the two-photon-subtracted squeezed state shows favorable
	displacement-fidelity radii over selected quadrature directions at matched
	\(\langle n\rangle\). These results identify a regime relevant to
	homodyne-aligned continuous-variable control and anisotropic
	displacement-noise mitigation, with directional sensing as a natural dual
	application.
\end{abstract}
	
	
	\maketitle
	
	\section{Introduction}
	
	Non-Gaussian optical states are essential resources for continuous-variable
	(CV) quantum information processing. Gaussian states and Gaussian operations
	alone admit efficient classical descriptions, while Wigner-function
	negativity provides a widely used diagnostic of genuinely non-Gaussian structure
	and a useful resource diagnostic for CV photonics
	\cite{Bartlett2002-wk,Kenfack2004-bg,Mari2012-hu,Veitch2012-ld,Walschaers2021-ld}.

Squeezed number states, displaced squeezed number states,
photon-subtracted squeezed states, squeezed-Fock generation protocols, and
optical Schr\"{o}dinger-cat states have all been studied extensively
\cite{Kim1989-ke,Moller1996-zo,Nieto1997-rt,Olivares2005-yw,
	Biswas2007-ni,Ourjoumtsev2006-dt,Neergaard-Nielsen2006-mm,
	Wakui2007-qj,Gerrits2010-vr,Takase2021-co,Korolev2023-vs,
	Bashmakova2025-tc}. Recently squeezed-Fock and
conditionally squeezed states have been proposed directly as bosonic-code resources
\cite{Bashmakova2025-lm,Hope2025-vl}.
The purpose of the present work is therefore not to introduce a new state
family, a new squeezed-Fock generation protocol, or a new bosonic code.
Rather, we use photon-conditioned squeezed states as benchmarked resources
and address how their matched-energy scalar and directional displacement metrics
compare with Fock and cat references.

The specific gap addressed here is the absence of a common matched-energy resource map that treats these established families on the same footing. Such a map separates scalar Wigner negativity from directional displacement response and clarifies how the same photon budget is distributed between non-Gaussian excitation and Gaussian squeezing. Our approach provides a common matched-\(\langle n\rangle\) baseline before platform-specific costs are included.
	
	Squeezed light provides a natural parent resource for this setting. A
	squeezed vacuum \(S(r,\theta)\ket{0}\) is Gaussian and Wigner-positive, but
	conditional photon subtraction converts it into a non-Gaussian state.
	In the weak-tap limit, a heralding event implements the annihilation
	operator on the transmitted mode. A single subtraction from squeezed vacuum
	prepares, up to normalization and phase, the squeezed single-photon state
	\(S(r,\theta)\ket{1}\). Higher-order subtraction produces parity-restricted
	squeezed Fock superpositions. In particular, two-photon subtraction does not
	generically prepare a pure \(S(r,\theta)\ket{2}\) state. This distinction is
	important for connecting ideal squeezed-Fock targets with experimentally
	produced photon-conditioned states.
	
	We develop a matched-mean-photon-number resource map for the sequence
	\begin{equation}
		S(r)\ket{0},\qquad
		aS(r)\ket{0},\qquad
		a^2S(r)\ket{0},
	\end{equation}
	and compare these states with Fock and coherent-cat benchmarks. The common
	resource budget is
	\begin{equation}
		\langle n\rangle=\langle a^\dagger a\rangle,
	\end{equation}
	which fixes the excitation energy above the vacuum for a single optical
	mode. We evaluate both scalar Wigner metrics,
	negativity, \(\delta\), and negativity/photon, \(\delta/\langle n\rangle\), and directional displacement fidelity 
	metrics based on
	\begin{equation}
		F_\psi(\alpha)
		=
		\left|
		\bra{\psi}D(\alpha)\ket{\psi}
		\right|^2 .
	\end{equation}
	
	The main result is a metric-dependent resource ordering. Scalar Wigner negativity and directional displacement-fidelity response do not rank the state families in the same order. Cat benchmarks can remain favorable under scalar negativity
	measures, while photon-conditioned squeezed states can be favorable under
	directional displacement-response measures when the noise or measurement
	axis is known. This metric-dependent ordering is the central identified resource tradeoff.
	
	The paper is organized as follows. Section~\ref{sec:theory} establishes the
	state definitions, photon-subtraction identities, and displacement-fidelity
	scaling. Section~\ref{sec:methods} describes the matched-\(\langle n\rangle\)
	benchmarking procedure. Section~\ref{sec:results} presents the state
	landscape, scalar Wigner metrics, and directional displacement-fidelity
	radii. Section~\ref{sec:discussion} discusses the implications for
	homodyne-aligned displacement-noise mitigation and sensing.
	
	\section{Theory}
	\label{sec:theory}
	
	\subsection{Conventions}
	
	We use
	\begin{equation}
		[a,a^\dagger]=1,\qquad
		x=\frac{a+a^\dagger}{\sqrt{2}},\qquad
		p=\frac{a-a^\dagger}{i\sqrt{2}},
	\end{equation}
	so that \([x,p]=i\). The Wigner function is normalized as
	\begin{equation}
		\int_{\mathbb{R}^2} W_\rho(x,p)\,dx\,dp=1.
	\end{equation}
	At the origin it is fixed by photon-number parity,
	\begin{equation}
		W_\rho(0,0)=\frac{1}{\pi}\mathrm{Tr}[\Pi\rho],
		\qquad
		\Pi=(-1)^{a^\dagger a}.
		\label{eq:parity_origin}
	\end{equation}
	Thus
	\begin{equation}
		W_{\ket{n}}(0,0)=\frac{(-1)^n}{\pi}.
	\end{equation}
	This provides a direct parity diagnostic for the states shown in
	Fig.~\ref{fig:resource_landscape}.
	
	The single-mode squeezing operator is
	\begin{equation}
		S(r,\theta)
		=
		\exp\!\left[
		\frac{1}{2}
		\left(
		r e^{-i\theta}a^2
		-
		r e^{i\theta}a^{\dagger 2}
		\right)
		\right].
		\label{eq:squeezer}
	\end{equation}
	Unless otherwise stated, the numerical examples use real squeezing,
	\(\theta=0\). Squeezing in decibels is related to \(r\) by
	\begin{equation}
		r = \frac{\ln 10}{20}\,r_{\mathrm{dB}}.
		\label{eq:db_to_r}
	\end{equation}
	
	\subsection{Benchmark state families}
	
	The Fock state is
	\begin{equation}
		\ket{n}=\frac{(a^\dagger)^n}{\sqrt{n!}}\ket{0},
	\end{equation}
	and a coherent state is
	\begin{equation}
		\ket{\alpha}=D(\alpha)\ket{0},
		\qquad
		D(\alpha)=\exp(\alpha a^\dagger-\alpha^*a).
	\end{equation}
	Coherent states are Gaussian and Wigner-positive; in this work they enter
	only as components of cat-state benchmarks.
	
	The even and odd cat states are
	\begin{equation}
		\ket{\mathcal C_\pm(\alpha)}
		=
		\mathcal N_\pm
		\left(
		\ket{\alpha}\pm\ket{-\alpha}
		\right).
	\end{equation}
	Their mean photon numbers are
	\begin{equation}
		\langle n\rangle_{\mathcal C_-}
		=
		|\alpha|^2\coth|\alpha|^2,
		\qquad
		\langle n\rangle_{\mathcal C_+}
		=
		|\alpha|^2\tanh|\alpha|^2 .
		\label{eq:cat_photon_numbers}
	\end{equation}
	At large \(|\alpha|\), the coherent components become nearly orthogonal
	because
	\begin{equation}
		\braket{\alpha|-\alpha}=e^{-2|\alpha|^2}.
	\end{equation}
	Consequently, even and odd cats become nearly degenerate in several
	displacement-response metrics at sufficiently large matched energy.
	
	For squeezed Fock states,
	\begin{equation}
		\ket{\psi_n(r,\theta)}=S(r,\theta)\ket{n},
	\end{equation}
	the mean photon number is
	\begin{equation}
		\langle n\rangle_{S\ket{n}}
		=
		n+(2n+1)\sinh^2 r.
		\label{eq:squeezed_fock_photon_number}
	\end{equation}
	In particular,
	\begin{equation}
		\langle n\rangle_{S\ket{0}}=\sinh^2 r,
		\qquad
		\langle n\rangle_{S\ket{1}}=1+3\sinh^2 r.
	\end{equation}
	
	\subsection{Photon subtraction from squeezed vacuum}
	
	Photon subtraction is modeled by the action of \(a\) on the squeezed state,
	as realized approximately by a weak tap and conditional heralding event.
	Using the Bogoliubov transform we obtain
	\begin{equation}
		\begin{gathered}
			S^\dagger(r,\theta)aS(r,\theta)
			=
			\mu a+\nu a^\dagger,\\
			\mu=\cosh r,\qquad
			\nu=-e^{i\theta}\sinh r.
		\end{gathered}
	\end{equation}
	and
	\begin{align}
		aS(r,\theta)\ket{0}
		&=
		S(r,\theta)
		\left(
		\mu a+\nu a^\dagger
		\right)\ket{0}
		\nonumber\\
		&=
		\nu\,S(r,\theta)\ket{1}.
		\label{eq:single_subtraction}
	\end{align}
	Thus the normalized one-photon-subtracted squeezed vacuum is exactly a
	squeezed single-photon state, up to an overall phase.
	
	For two-photon subtraction,
	\begin{align}
		a^2S(r,\theta)\ket{0}
		&=
		S(r,\theta)
		\left(
		\mu a+\nu a^\dagger
		\right)^2
		\ket{0}
		\nonumber\\
		&=
		S(r,\theta)
		\left(
		\mu\nu\ket{0}
		+
		\sqrt{2}\nu^2\ket{2}
		\right).
		\label{eq:double_subtraction}
	\end{align}
	The two-click state is therefore an even-parity squeezed superposition, not
	a pure \(S(r,\theta)\ket{2}\) state. The relative amplitude of the
	\(\ket{2}\) and \(\ket{0}\) components in the squeezed frame is
	\begin{equation}
		\frac{\sqrt{2}\nu^2}{\mu\nu}
		=
		\sqrt{2}\frac{\nu}{\mu}
		=
		-\sqrt{2}e^{i\theta}\tanh r .
		\label{eq:double_ratio}
	\end{equation}
	Thus, in the large-squeezing limit, the normalized state in the squeezed
	frame approaches an even superposition proportional to
	\begin{equation}
		\ket{0}-\sqrt{2}e^{i\theta}\ket{2},
		\label{eq:chi_infty}
	\end{equation}
	rather than a pure two-photon state. This provides a useful analytic
	reference for the numerical resource curves.
	
	\subsection{Scalar Wigner negativity}
	
	We quantify Wigner negativity, \(\delta\), using
	\begin{equation}
		\delta
		=
		\frac{1}{2}
		\left[
		\int_{\mathbb{R}^2}|W(x,p)|\,dx\,dp
		-1
		\right].
		\label{eq:wigner_negativity}
	\end{equation}
	For ideal Gaussian unitary transformations, including squeezing and phase
	rotations, \(\delta\) is invariant under the corresponding symplectic
	coordinate transformation. Squeezing therefore does not create additional
	integrated Wigner negativity. Instead, it redistributes existing
	non-Gaussian structure in phase space. This observation is important for
	interpreting the comparison with cat states: a squeezed single-photon state
	can have the same \(\delta\) as \(\ket{1}\) while exhibiting a very different
	directional displacement response.
	
	\subsection{Displacement-fidelity radius}
	
	For a pure state \(\ket{\psi}\), we define the displacement fidelity
	\begin{equation}
		F_\psi(\alpha)
		=
		\left|
		\bra{\psi}D(\alpha)\ket{\psi}
		\right|^2.
		\label{eq:displacement_fidelity}
	\end{equation}

This overlap is the squared magnitude of the symmetrically ordered
characteristic function evaluated at the displacement amplitude. Up to the usual convention-dependent scaling of the complex phase-space
argument, it is the same object that appears in displaced-number-state and
displaced-squeezed-number-state overlap formulas. It therefore
directly probes how distinguishable the displaced state
\(D(\alpha)\ket{\psi}\) is from the original state \(\ket{\psi}\). In the
present work we use this quantity in two complementary ways. A slowly
decaying fidelity indicates robustness against displacement noise, whereas
a rapidly decaying fidelity indicates sensitivity to weak displacement
signals. 
In the numerical results, \(\epsilon\) denotes the complex displacement
amplitude in \(D(\epsilon e^{i\phi})\), not the physical quadrature
translation distance.	
	For a displacement direction \(\phi\), the fidelity-threshold radius is
	\begin{equation}
		R_F(\phi)
		=
		\max\left\{
		\epsilon:
		F_\psi(\epsilon e^{i\phi})\ge F_{\rm th}
		\right\}.
		\label{eq:radius_phi}
	\end{equation}

	This is not a quantum-error-correction distance; no recovery map is assumed.
	It is a direct displacement-response diagnostic. Larger \(R_F\) means the
	state remains close to itself under a larger coherent displacement in that
	direction.
	
For small displacements, the fidelity has a variance-controlled expansion.
A real displacement \(D(\epsilon)\) translates the state along one
phase-space quadrature and is generated by the conjugate quadrature; an
imaginary displacement \(D(i\epsilon)\) translates along the orthogonal
quadrature. Thus
\begin{align}
	1-F_x(\epsilon)
	&=
	A\,\mathrm{Var}(p)\,\epsilon^2
	+O(\epsilon^4),
	\label{eq:small_disp_x}\\
	1-F_p(\epsilon)
	&=
	A\,\mathrm{Var}(x)\,\epsilon^2
	+O(\epsilon^4),
	\label{eq:small_disp_p}
\end{align}
The proportionality constant \(A\) depends on whether \(\epsilon\) denotes the
complex displacement amplitude \(\alpha\) or the corresponding quadrature
translation.

The important point is independent of this convention: the two leading slopes are controlled by the conjugate quadrature variances. In the numerical
implementation, we verify this relation by fitting \(1-F\) versus
\(\epsilon^2\) at small \(\epsilon\) and comparing the fitted slopes with the corresponding quadrature variances.
	
	For a squeezed Fock state \(S(r)\ket{n}\), the quadrature variances scale as
	\begin{equation}
		\mathrm{Var}(x)\propto \left(n+\frac{1}{2}\right)e^{-2r},
		\qquad
		\mathrm{Var}(p)\propto \left(n+\frac{1}{2}\right)e^{+2r},
		\label{eq:variance_scaling}
	\end{equation}
	up to the sign convention for the squeeze axis. Therefore squeezing
	redistributes displacement response between conjugate quadratures. The
	corresponding threshold radii scale approximately as
	\begin{equation}
		R_x\sim e^{-r},
		\qquad
		R_p\sim e^{+r},
		\label{eq:radius_scaling}
	\end{equation}
	again up to convention and state-dependent constants. This analytic scaling
	is the basis for interpreting the directional-radius results below.
	
	\section{Numerical methods}
	\label{sec:methods}
	
	All comparisons are performed at matched mean photon number
	\(\langle n\rangle\). For each target value of \(\langle n\rangle\), we solve
	for the squeezing \(r\) or cat amplitude \(\alpha\) required to match that
	target using Eqs.~\eqref{eq:cat_photon_numbers} and
	\eqref{eq:squeezed_fock_photon_number}. 

The numerical parameters were kept fixed across state families within each
comparison. The baseline calculations used a Fock-space cutoff of 80 and a
uniform phase-space grid with 201 points along each quadrature direction over
the range \(x,p\in[-7,7]\). Numerical convergence was checked by repeating
the calculation with an increased Fock cutoff, an increased phase-space grid
density, and an enlarged phase-space window. Across these tests, the scalar Wigner metrics changed by less than
\(6\times 10^{-3}\) in \(\delta\) and \(2\times 10^{-3}\) in
\(\delta/\langle n\rangle\). The displacement-fidelity radii were unchanged
within numerical interpolation precision, and the ordering of state families
was unchanged. These checks indicate that the reported trends are numerical features of the matched-state resource map rather than artifacts of the chosen cutoff, grid density, or phase-space window.

The squeezing was constrained to
\(r_{\mathrm{dB}}\le 12.5\), corresponding to
\(r\le(\ln 10/20)12.5\), to remain within a near-term experimentally
relevant squeezing range. For the cat benchmarks, the defining real
\(\alpha\) values were obtained by matching the same mean photon-number
budget as the photon-conditioned squeezed states.

	Unless otherwise stated, we use \(F_{\rm th}=0.90\) across all state families, so that differences in
	\(R_x\), \(R_p\), and \(R_{\max}/R_{\min}\) reflect state response rather than a change in the fidelity criterion.
	
	The scalar comparison uses \(\delta\) and
	\(\delta/\langle n\rangle\). The directional comparison uses the threshold
	radii \(R_x=R_F(0)\), \(R_p=R_F(\pi/2)\), and the anisotropy ratio
	\begin{equation}
		\frac{R_{\max}}{R_{\min}}
		=
		\frac{\max_\phi R_F(\phi)}{\min_\phi R_F(\phi)}.
	\end{equation}
	In the main text, \(R_x\) and \(R_p\) are evaluated along the laboratory
	quadrature axes. Changing the squeezing phase rotates the radius contour;
	therefore the physically relevant angle is the relative orientation between
	the state and the homodyne local oscillator.
	
	The simulations include several internal consistency checks. First, the
	normalized state \(aS(r)\ket{0}\) is verified to coincide with
	\(S(r)\ket{1}\). Second, the Wigner negativity of \(S(r)\ket{1}\) is
	constant with squeezing parameter \(r\), as expected from Gaussian-unitary invariance. Third,
	unsqueezed Fock states provide isotropic displacement-radius references,
	with \(R_x=R_p\) and \(R_{\max}/R_{\min}=1\). Fourth, even and odd cat
	responses converge at larger \(\langle n\rangle\), consistent with the
	exponential suppression of \(\braket{\alpha|-\alpha}\). The displacement radii are extracted by
	linear interpolation at the first crossing of the chosen fidelity threshold.
	If no crossing occurs within the scanned displacement interval, the reported
	radius is treated as a lower bound. These checks are used to distinguish
	physical saturation or convergence from finite-window numerical artifacts.
	
	\section{Results}
	\label{sec:results}
	
	\subsection{State landscape}
	
\begin{figure*}[t]
		\centering
		\includegraphics[width=\textwidth]{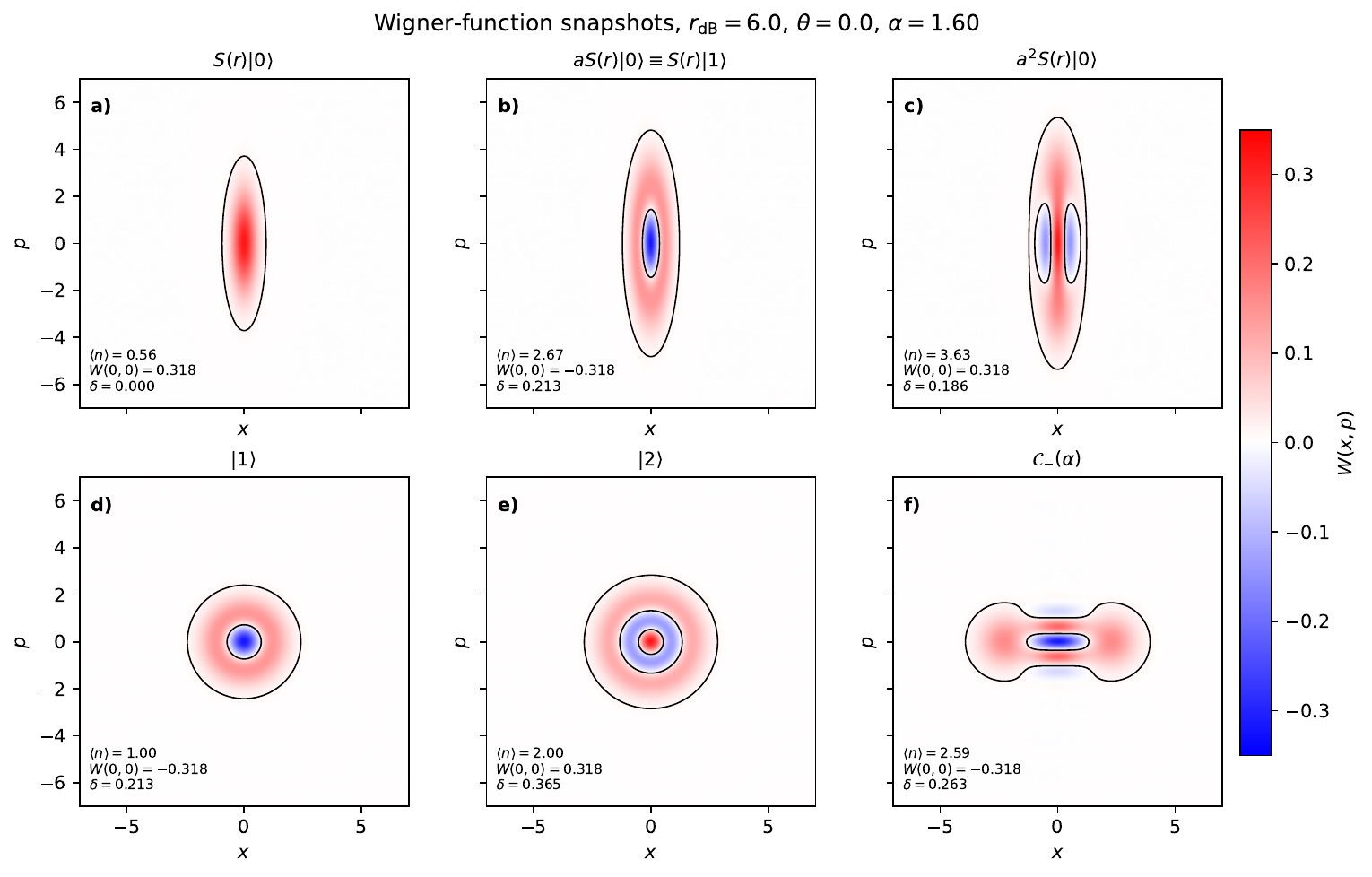}
		\caption{
Wigner-function snapshots of the state families used in the matched-energy
resource comparison. The top row follows the photon-conditioning route:
(a) squeezed vacuum \(S(r)\ket{0}\), the Gaussian parent state;
(b) the one-photon-subtracted state \(aS(r)\ket{0}\), which is proportional
to \(S(r)\ket{1}\) in the ideal limit; and
(c) the two-photon-subtracted state \(a^{2}S(r)\ket{0}\), an even-parity
squeezed Fock superposition. The bottom row shows canonical non-Gaussian
references:
(d) the odd Fock state \(\ket{1}\);
(e) the even Fock state \(\ket{2}\); and
(f) an odd coherent cat state
\(\mathrm{Cat}_{-}(\alpha)\propto\ket{\alpha}-\ket{-\alpha}\).
For the representative panels shown here, the squeezed-state snapshots use
\(r_{\mathrm{dB}}=6.0\) with \(\theta=0\), and the odd-cat benchmark uses
\(\alpha=1.6\).
All panels use the \([x,p]=i\) convention, a common phase-space window,
and the same Wigner color scale. The annotated values report the mean
photon number \(\langle n\rangle\), the parity diagnostic \(W(0,0)\), and
the integrated Wigner negativity \(\delta\).
		}
		\label{fig:resource_landscape}
	\end{figure*}

	Figure~\ref{fig:resource_landscape} summarizes the state families used in
	this work. The top row follows the experimentally motivated route from
	squeezed light to photon-conditioned non-Gaussian states. The squeezed
	vacuum \(S(r)\ket{0}\) in Fig.~\ref{fig:resource_landscape}(a) is the
	Gaussian parent resource. It is zero-displacement and anisotropic in phase
	space, but its Wigner function is nonnegative and therefore
	\(\delta=0\). Its role is not to compete as a non-Gaussian state, but to
	serve as the optical input from which the conditioned states are generated.
	
	A single subtraction event produces the state shown in
	Fig.~\ref{fig:resource_landscape}(b). As derived in
	Eq.~\eqref{eq:single_subtraction}, this state is proportional to
	\(S(r)\ket{1}\) in the ideal weak-tap limit. It has odd parity and therefore
	a negative Wigner value at the origin, consistent with
	Eq.~\eqref{eq:parity_origin}. The one-click state is therefore the most
	direct photon-conditioned representative of the squeezed-Fock ladder.
	
	The two-click state in Fig.~\ref{fig:resource_landscape}(c) shows the next
	member of the photon-conditioned sequence. Equation~\eqref{eq:double_subtraction}
	shows that this state is not simply \(S(r)\ket{2}\), but a squeezed
	superposition of \(\ket{0}\) and \(\ket{2}\). It has even parity and a
	positive central Wigner value while retaining non-Gaussian structure away
	from the origin. This panel is therefore important for distinguishing the
	ideal squeezed-Fock basis from the states produced by direct multi-photon
	subtraction.
	
	The bottom row provides reference states. The Fock states
	\(\ket{1}\) and \(\ket{2}\) isolate the odd- and even-parity Wigner
	structures before squeezing redistributes them in phase space. The odd cat
	state provides a displaced non-Gaussian benchmark. Overall, photon subtraction converts a Gaussian squeezed parent into
	zero-displacement, parity-resolved non-Gaussian states.
	
	\subsection{Matched-energy scalar resource metrics}
	
\begin{figure}[h]
		\centering
		\includegraphics[width=0.75\textwidth,height=0.7\textheight,keepaspectratio]{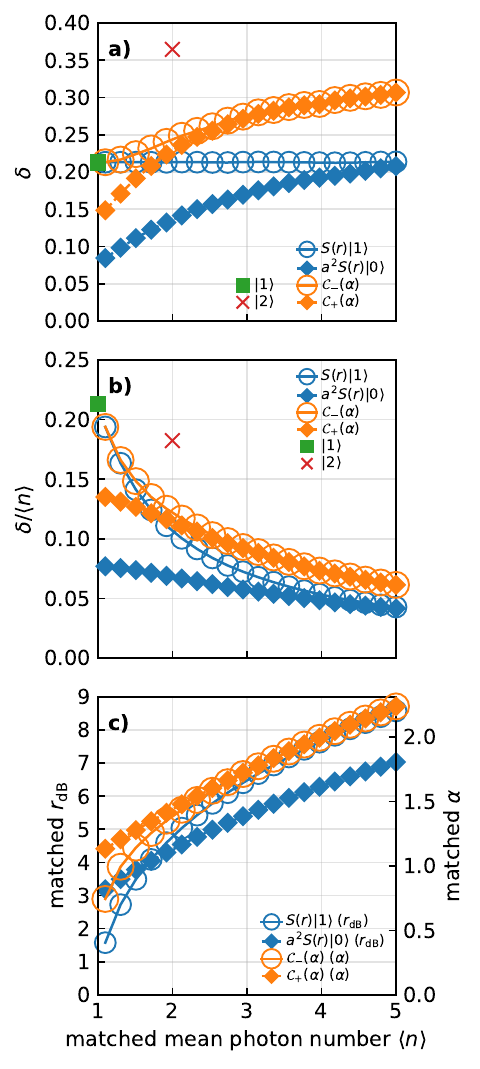}
		\caption{
			Matched-\(\langle n\rangle\) scalar resource metrics. 
			(a) Integrated Wigner negativity \(\delta\) versus matched mean photon
			number. The Fock states \(\ket{1}\) and \(\ket{2}\) are included as
			marker-only references. 
			(b) Energy-normalized negativity \(\delta/\langle n\rangle\). 
			(c) Matched parameter cost: squeezing \(r_{\rm dB}\) for the
			photon-conditioned squeezed families and cat amplitude \(\alpha\) for
			the cat benchmarks.
		}
		\label{fig:equal_energy}
	\end{figure}

	Figure~\ref{fig:equal_energy} shows the scalar resource metrics at matched
	mean photon number. For each point, \(r\) or \(\alpha\) is chosen so that
	the state has the specified target \(\langle n\rangle\). This removes a
	major ambiguity in visual Wigner-function comparisons: larger phase-space
	extent does not necessarily imply more non-Gaussianity per photon.
	
	Figure~\ref{fig:equal_energy}(a) shows that the squeezed single-photon state
	does not gain additional integrated negativity as its mean photon number is
	increased by squeezing. This is expected: \(S(r)\ket{1}\) is related to
	\(\ket{1}\) by a Gaussian unitary, so the total Wigner negativity is
	preserved. The behavior of \(a^2S(r)\ket{0}\) is different because its
	normalized state in the squeezed frame changes with \(r\). At large
	squeezing it approaches the limiting even superposition described by
	Eq.~\eqref{eq:chi_infty}, rather than a pure \(S(r)\ket{2}\) state.
	
	Figure~\ref{fig:equal_energy}(b) reports the energy-normalized quantity
	\(\delta/\langle n\rangle\). This panel emphasizes that extra squeezing
	increases photon number without necessarily increasing total integrated
	negativity. Cat-state benchmarks remain strong in scalar negativity metrics
	because the coherent separation controls the interference structure that
	generates Wigner negativity. Thus, the photon-conditioned squeezed states
	should not be interpreted as universally superior scalar non-Gaussian
	resources. 
	Their advantage, if present for a given task, must come from a
	different property. This conclusion is useful because it prevents an overinterpretation of the
	squeezed-Fock route. Photon conditioning does not provide a universal
	increase in scalar Wigner negativity per photon. Instead, it provides an
	experimentally natural, origin-centered way to introduce parity-resolved
	non-Gaussianity into a squeezed optical mode. The question is therefore
	whether this state structure is advantageous for a task whose figure of
	merit is not simply the total negative Wigner volume.
	
	Figure~\ref{fig:equal_energy}(c) shows the matched parameter cost. It
	translates the equal-energy comparison into experiment: squeezing
	for the photon-conditioned squeezed states and coherent amplitude for the
	cat benchmarks. This panel is not itself a resource metric, but it makes
	the comparison operational.
	
	\subsection{Directional displacement-fidelity radii}
	
\begin{figure}[!htbp]
		\centering
		\includegraphics[width=0.75\textwidth,height=0.7\textheight,keepaspectratio]{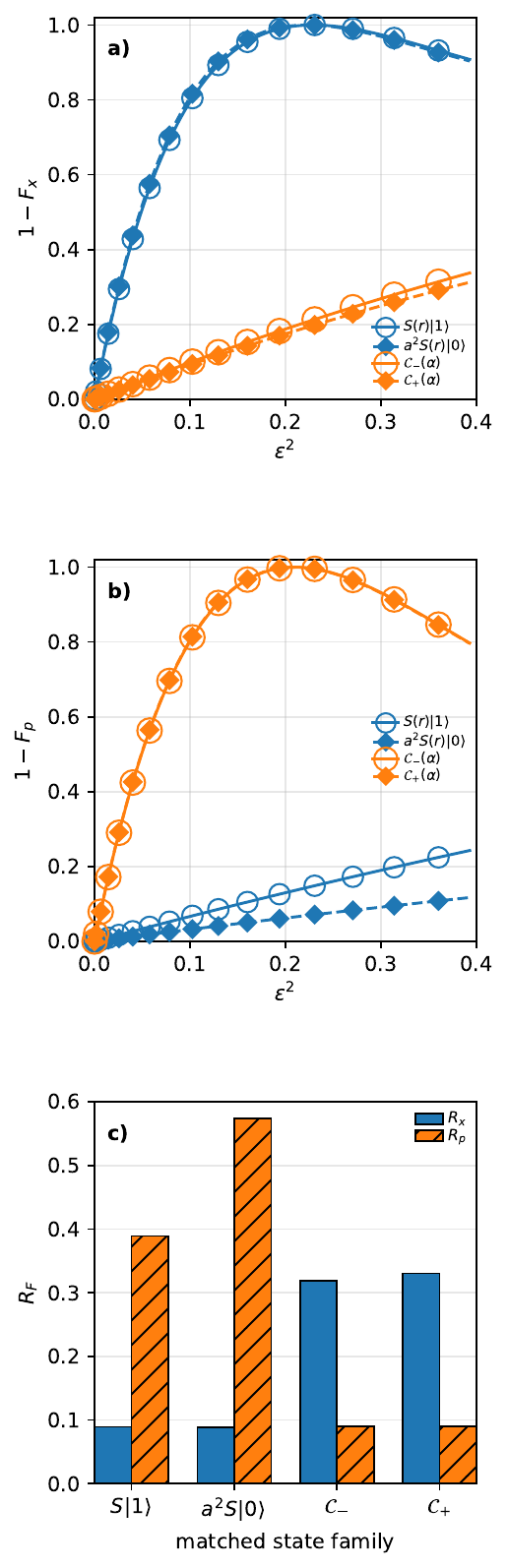}
		\caption{
			Matched-\(\langle n\rangle\) displacement-fidelity radii. 
			(a) Radius \(R_x\) for displacements along the \(x\) quadrature.
			(b) Radius \(R_p\) for displacements along the \(p\) quadrature.
			(c) Anisotropy ratio \(R_{\max}/R_{\min}\). Fock states are included as
			isotropic marker-only references, for which \(R_x=R_p\) and
			\(R_{\max}/R_{\min}=1\). The radius is a fidelity-threshold diagnostic,
			not a full quantum-error-correction distance.
		}
		\label{fig:displacement_radius}
	\end{figure}

	We next evaluate directional displacement response using the same
	matched-\(\langle n\rangle\) states. Figure~\ref{fig:displacement_radius}
	shows \(R_x\), \(R_p\), and the anisotropy ratio \(R_{\max}/R_{\min}\) at a
	fixed fidelity threshold. Larger \(R\) means the state remains close to
	itself under a larger coherent displacement in that direction.
	
	The displacement-radius comparison reveals a tradeoff that is not visible
	in the scalar negativity metrics. Along one quadrature, photon-conditioned
	squeezed states can exhibit a substantially larger radius than the Fock and
	cat references over the matched-energy range shown. Along the orthogonal
	quadrature they can be more sensitive. The two-photon-subtracted state
	\(a^2S(r)\ket{0}\) is especially anisotropic: it can have the largest
	favorable-axis radius while also having the largest
	\(R_{\max}/R_{\min}\). This should not be interpreted as a uniform advantage over cat benchmarks:
	the same state is less robust along the orthogonal quadrature. Its advantage
	is directional and therefore task dependent.
	
	This behavior is the expected signature of squeezing. Since ideal squeezing
	does not increase integrated Wigner negativity, its effect appears instead
	as a redistribution of displacement response between conjugate quadratures.
	The squeezed-state advantage is therefore directional rather than universal.
	It is useful when the dominant displacement noise is anisotropic or can be
	aligned with the favorable quadrature by phase control. In this sense, Fig.~\ref{fig:displacement_radius} provides the operational
	counterpart to Fig.~\ref{fig:equal_energy}. The scalar metrics show how much
	Wigner negativity is present at a fixed photon budget, while the radius
	metrics show where that non-Gaussian structure is tolerant or sensitive to
	coherent shifts. The relevant resource is therefore not simply a larger
	\(\delta\), but the ability to orient the displacement response relative to
	a known noise direction or homodyne measurement axis.
	The representative polar contour in Appendix~\ref{app:polar} confirms that
	this is not a single-axis artifact: the favorable response of the
	two-photon-subtracted squeezed state extends over a finite angular sector
	around the stretched quadrature. The axis-specific radii in
	Fig.~\ref{fig:displacement_radius} should therefore be interpreted as cuts
	through a full directional response surface \(R_F(\phi)\).
	The cat benchmarks show a different angular mechanism. For real \(\alpha\),
	the cat components are separated along one phase-space direction. At larger
	matched \(\langle n\rangle\), the overlap
	\(\braket{\alpha|-\alpha}\) becomes small, and the even and odd cat
	displacement responses become nearly identical. This convergence provides
	an additional sanity check on the simulation and emphasizes that both cat
	states and squeezed-Fock resources can be angularly tunable, but by
	different physical mechanisms: coherent separation for cats and squeezing
	axis control for photon-conditioned squeezed states.
	
	\section{Discussion}
	\label{sec:discussion}
	
	\subsection{Homodyne-aligned displacement-noise mitigation}
	
The main application suggested by these results is homodyne-aligned
anisotropic displacement-noise mitigation. Homodyne detection is a standard
phase-sensitive measurement in CV photonics and is central to both optical
non-Gaussian state characterization and measurement-based CV protocols
\cite{Lenzini2018-su,Eaton2022-ed}.
	
	In a phase-referenced CV optical
	system, a local oscillator defines the measured quadrature. If the dominant
	displacement noise is known or can be aligned relative to this quadrature,
	then the relevant resource metric is not an angle-averaged scalar
	non-Gaussianity, but the displacement-fidelity radius in the noise
	direction.
	
	This is the regime where photon-conditioned squeezed states can be useful
	despite not universally maximizing \(\delta\) or
	\(\delta/\langle n\rangle\). Their robust axis is controlled by the
	squeezing phase and remains centered at the origin. Cat states also have a
	directional response, but their anisotropy is tied to coherent-state
	separation. Thus the comparison distinguishes two mechanisms of angular
	response rather than identifying a unique capability of either family.
	
The same anisotropy has a dual interpretation for sensing. Along the
large-radius axis, a state is robust against displacement noise. Along the
small-radius axis, the local slope
\begin{equation}
	\Gamma(\phi)
	=
	\left.
	\frac{d}{d\epsilon^2}
	\left[
	1-F_\psi(\epsilon e^{i\phi})
	\right]
	\right|_{\epsilon=0}
\end{equation}
is large, which indicates strong sensitivity to weak displacements in that
direction with expected applications to sensing.
	
\subsection{Relation to known state families and recent code proposals}

The state families studied here are established. Squeezed number states and
displaced squeezed number states have long been analyzed, including inner
products relevant to displacement overlaps
\cite{Kim1989-ke,Moller1996-zo,Nieto1997-rt}. Photon-subtracted squeezed
states and optical cat states have also been studied both theoretically and
experimentally
\cite{Olivares2005-yw,Biswas2007-ni,Ourjoumtsev2006-dt,
	Neergaard-Nielsen2006-mm,Wakui2007-qj,Gerrits2010-vr,Takase2021-co}.
Recent measurement-based protocols further address the generation of
squeezed Fock states from Gaussian resources with photon-number conditioning
\cite{Korolev2023-vs}.

Squeezed-Fock states have recently been considered as bosonic-code
resources and compared with squeezed-cat codewords for particle-loss and
dephasing channels \cite{Bashmakova2025-tc,Bashmakova2025-lm}. Related
conditionally squeezed oscillator states have been proposed in
qubit-oscillator platforms with applications to error mitigation
\cite{Hope2026-ry}. These works address generation protocols, encoded
codewords, and channel-specific recovery or error-mitigation performance.

Our comparison is positioned differently. We do not define logical codewords, optimize a recovery map, or propose a new generation scheme.
Instead, we use matched mean photon number to compare photon-conditioned
squeezed states, Fock references, and cat benchmarks at the state level. The
comparison separates scalar Wigner negativity from directional
displacement-fidelity response, which is the metric relevant to
homodyne-aligned displacement-noise geometry. This distinction matters
because different tasks favor different resources: cat states can be strong
scalar Wigner-negativity resources; Fock states provide isotropic
displacement-response references; and photon-conditioned squeezed states
provide origin-centered non-Gaussian resources whose directional response is
set by the squeeze axis.

\subsection{Resource tradeoff}

The comparison suggests a simple division of roles. Cat states are strong
benchmarks for scalar Wigner negativity and coherent-superposition
interference. Fock states provide isotropic non-Gaussian references with no
preferred displacement direction. Photon-conditioned squeezed states occupy
a different regime: they are origin-centered, parity-resolved, and
anisotropic. They need not maximize \(\delta\) or
\(\delta/\langle n\rangle\) in order to be useful; their relevant advantage
appears when the task has a preferred phase-space direction.

This tradeoff can be summarized as follows. If the task rewards total
negative Wigner volume at fixed photon number, cat benchmarks are natural
competitors. If the task requires an isotropic displacement response, Fock
states provide the clean reference. If the task has a known quadrature
geometry, photon-conditioned squeezed states offer a tunable response:
the squeeze phase can align either the robust axis or the sensitive axis
with the experimentally relevant direction. This is the regime targeted by
the present work.

	\subsection{Limitations and outlook}

This work does not include optical loss, finite detector efficiency, mode
mismatch, or finite heralding probability. These effects are essential for a
complete experimental performance model of photon-subtracted squeezed states
and optical cat generation
\cite{Ourjoumtsev2006-dt,Neergaard-Nielsen2006-mm,Wakui2007-qj,Gerrits2010-vr,Takase2021-co}. The
	present analysis also does not define a recovery operation, so the
	displacement-fidelity radius should not be interpreted as a full
	error-correction distance.
	
	More broadly, the same phase-space benchmarking strategy can serve as a diagnostic for assessing the value of state-generation targets before substantial experimental effort is invested. This may be especially useful for structured bosonic states proposed for error-correction or protected-information applications, where preparation complexity can be high and utility is not captured by photon number or squeezing alone. By comparing Wigner negativity, phase-space structure, and characteristic length scales prior to implementation, one can identify which target states are most likely to provide useful nonclassical resources and which may offer limited practical advantage despite their preparation complexity.
	
	The connection to grid-state synthesis is left for future work.
	The GKP code protects against phase-space shift errors
	\cite{Gottesman2001-ke}, and cat-state breeding provides an established route
	toward approximate grid states using linear optics and homodyne measurements
	\cite{Weigand2018-qe}. Recent optical schemes have also shown how squeezed cat
	and GKP states can be prepared deterministically from photon-number
	resources \cite{Winnel2024-om}. A single squeezed-Fock or
	photon-conditioned squeezed state is not a grid state, because it lacks the
	translational comb structure required for a GKP codeword. We do not evaluate
	grid-state fidelity or a breeding protocol here. Nevertheless, the
	origin-centered, parity-resolved Hermite-Gaussian structure of
	photon-conditioned squeezed states may be useful as a seed in 
	conditional grid-state generation protocols.

\section{Conclusion}

We have analyzed photon-conditioned squeezed light as an origin-centered
non-Gaussian resource for directional displacement response in
continuous-variable photonics. Single-photon subtraction from squeezed
vacuum prepares a state proportional to \(S(r)\ket{1}\), while two-photon
subtraction prepares an even squeezed Fock superposition rather than a pure
\(S(r)\ket{2}\). This algebraic distinction is essential for interpreting
the resource metrics.

At matched mean photon number, cat benchmarks remain strong in scalar
Wigner-negativity metrics. Photon-conditioned squeezed states do not
dominate \(\delta\) or \(\delta/\langle n\rangle\). Their advantage appears
instead in directional displacement response: squeezing redistributes
fidelity radii between conjugate quadratures, allowing favorable-axis
robustness at the cost of orthogonal sensitivity. This tradeoff is naturally
suited to homodyne-aligned CV settings in which the dominant
displacement-noise direction is known or controllable.

The broader message is that non-Gaussian resources should not be ranked by a
single scalar metric alone. Matched-\(\langle n\rangle\) resource mapping
shows that photon-conditioned squeezed states, Fock states, and cat states
occupy different regions of a tradeoff space involving Wigner negativity,
photon-number cost, and directional displacement response. In this tradeoff
space, the value of photon-conditioned squeezed states is not universal
dominance, but origin-centered non-Gaussianity with a squeeze-axis-controlled
response that can be matched to anisotropic displacement-noise geometry.
	
\begin{acknowledgements}
	B.K.\ gratefully acknowledges support from the U.S. National Science Foundation through QCAP-Pilot, award~OSI-2410813. O.P.\ and B.K.\ acknowledge support from the U.S. National Science Foundation through the QCAP-Design program, award~OSI-2531569.
\end{acknowledgements}

\section*{Data and code availability}
The Python code used to generate the numerical results is available at
\url{https://github.com/boriskiefer/sim_squeezed_Fock}. The repository
contains the research code for constructing the matched-energy state families,
computing Wigner negativity, evaluating displacement-fidelity radii, and
reproducing the manuscript figures from the numerical workflow. Generated data
and figure files are not stored in the repository and can be regenerated by
running the provided code.

	\appendix
	
	\section{Representative angular displacement-radius contour}
	\label{app:polar}
	
\begin{figure}[h]
		\centering
		\includegraphics[width=0.75\textwidth,height=0.7\textheight,keepaspectratio]{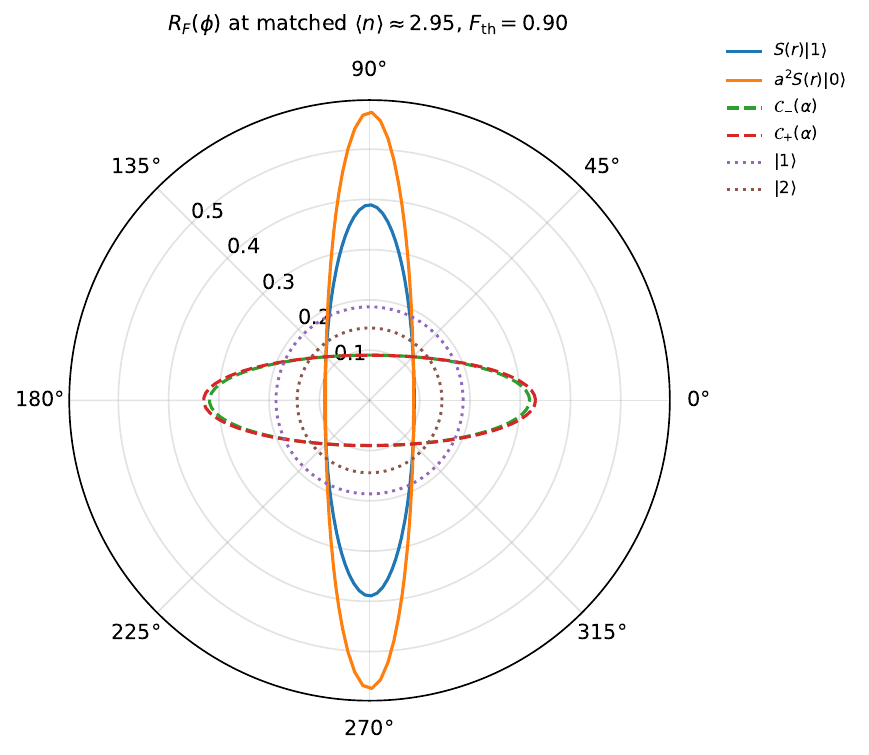}
		\caption{
			Angular displacement-fidelity radius \(R_F(\phi)\) at matched
			\(\langle n\rangle\simeq 3\). The Fock-state references are isotropic,
			while cat and photon-conditioned squeezed states show anisotropic
			displacement response. The two-photon-subtracted squeezed state displays
			an enlarged favorable-axis radius over a finite angular sector, whereas
			the even and odd cat contours are nearly identical at this energy because
			\(\braket{\alpha|-\alpha}\) is already small.
		}
		\label{fig:si_polar}
	\end{figure}

	The main text reports axis-specific radii \(R_x\) and \(R_p\) over a range
	of matched mean photon numbers. To verify that these are not single-axis
	artifacts, Fig.~\ref{fig:si_polar} shows the full angular radius
	\(R_F(\phi)\) at a representative matched value
	\(\langle n\rangle\simeq 3\). This value lies near the middle of the range
	used in Figs.~\ref{fig:equal_energy} and \ref{fig:displacement_radius}.
	
The polar contour also provides an angular-sector interpretation. For a
chosen reference state \(\ket{\psi_{\rm ref}}\), one can define
\begin{equation}
	\Omega_{\rm adv}
	=
	\left\{
	\phi:
	R_{\rm test}(\phi)>R_{\rm ref}(\phi)
	\right\}.
\end{equation}
The angular measure of \(\Omega_{\rm adv}\) identifies directions of
displacement noise for which the tested state provides a larger
fidelity-threshold radius than the reference.

For comparisons with cat benchmarks, one may also define a tolerance-based
sector,
\begin{equation}
	\Omega_{\eta}
	=
	\left\{
	\phi:
	R_{\rm test}(\phi)\ge \eta R_{\rm cat}(\phi)
	\right\},
\end{equation}
with \(0<\eta<1\). This identifies angular directions over which a
photon-conditioned squeezed state is comparable to a cat benchmark rather
than strictly larger. This sector-based view is the natural extension of the
axis-specific radii reported in the main text.
	
	\bibliographystyle{apsrev4-2}
	\bibliography{squeezed_Fock_2026}

\begin{thebibliography}{25}%
\makeatletter
\providecommand \@ifxundefined [1]{%
 \@ifx{#1\undefined}
}%
\providecommand \@ifnum [1]{%
 \ifnum #1\expandafter \@firstoftwo
 \else \expandafter \@secondoftwo
 \fi
}%
\providecommand \@ifx [1]{%
 \ifx #1\expandafter \@firstoftwo
 \else \expandafter \@secondoftwo
 \fi
}%
\providecommand \natexlab [1]{#1}%
\providecommand \enquote  [1]{``#1''}%
\providecommand \bibnamefont  [1]{#1}%
\providecommand \bibfnamefont [1]{#1}%
\providecommand \citenamefont [1]{#1}%
\providecommand \href@noop [0]{\@secondoftwo}%
\providecommand \href [0]{\begingroup \@sanitize@url \@href}%
\providecommand \@href[1]{\@@startlink{#1}\@@href}%
\providecommand \@@href[1]{\endgroup#1\@@endlink}%
\providecommand \@sanitize@url [0]{\catcode `\\12\catcode `\$12\catcode
  `\&12\catcode `\#12\catcode `\^12\catcode `\_12\catcode `\%12\relax}%
\providecommand \@@startlink[1]{}%
\providecommand \@@endlink[0]{}%
\providecommand \url  [0]{\begingroup\@sanitize@url \@url }%
\providecommand \@url [1]{\endgroup\@href {#1}{\urlprefix }}%
\providecommand \urlprefix  [0]{URL }%
\providecommand \Eprint [0]{\href }%
\providecommand \doibase [0]{https://doi.org/}%
\providecommand \selectlanguage [0]{\@gobble}%
\providecommand \bibinfo  [0]{\@secondoftwo}%
\providecommand \bibfield  [0]{\@secondoftwo}%
\providecommand \translation [1]{[#1]}%
\providecommand \BibitemOpen [0]{}%
\providecommand \bibitemStop [0]{}%
\providecommand \bibitemNoStop [0]{.\EOS\space}%
\providecommand \EOS [0]{\spacefactor3000\relax}%
\providecommand \BibitemShut  [1]{\csname bibitem#1\endcsname}%
\let\auto@bib@innerbib\@empty
\bibitem [{\citenamefont {Bartlett}\ \emph {et~al.}(2002)\citenamefont
  {Bartlett}, \citenamefont {Sanders}, \citenamefont {Braunstein},\ and\
  \citenamefont {Nemoto}}]{Bartlett2002-wk}%
  \BibitemOpen
  \bibfield  {author} {\bibinfo {author} {\bibfnamefont {S.~D.}\ \bibnamefont
  {Bartlett}}, \bibinfo {author} {\bibfnamefont {B.~C.}\ \bibnamefont
  {Sanders}}, \bibinfo {author} {\bibfnamefont {S.~L.}\ \bibnamefont
  {Braunstein}},\ and\ \bibinfo {author} {\bibfnamefont {K.}~\bibnamefont
  {Nemoto}},\ }\href@noop {} {\bibfield  {journal} {\bibinfo  {journal} {Phys.
  Rev. Lett.}\ }\textbf {\bibinfo {volume} {88}},\ \bibinfo {pages} {097904}
  (\bibinfo {year} {2002})}\BibitemShut {NoStop}%
\bibitem [{\citenamefont {Kenfack}\ and\ \citenamefont
  {Yczkowski}(2004)}]{Kenfack2004-bg}%
  \BibitemOpen
  \bibfield  {author} {\bibinfo {author} {\bibfnamefont {A.}~\bibnamefont
  {Kenfack}}\ and\ \bibinfo {author} {\bibfnamefont {K.}~\bibnamefont
  {Yczkowski}},\ }\href@noop {} {\bibfield  {journal} {\bibinfo  {journal} {J.
  Opt. B Quantum Semiclassical Opt.}\ }\textbf {\bibinfo {volume} {6}},\
  \bibinfo {pages} {396} (\bibinfo {year} {2004})}\BibitemShut {NoStop}%
\bibitem [{\citenamefont {Mari}\ and\ \citenamefont
  {Eisert}(2012)}]{Mari2012-hu}%
  \BibitemOpen
  \bibfield  {author} {\bibinfo {author} {\bibfnamefont {A.}~\bibnamefont
  {Mari}}\ and\ \bibinfo {author} {\bibfnamefont {J.}~\bibnamefont {Eisert}},\
  }\href@noop {} {\bibfield  {journal} {\bibinfo  {journal} {Phys. Rev. Lett.}\
  }\textbf {\bibinfo {volume} {109}},\ \bibinfo {pages} {230503} (\bibinfo
  {year} {2012})}\BibitemShut {NoStop}%
\bibitem [{\citenamefont {Veitch}\ \emph {et~al.}(2012)\citenamefont {Veitch},
  \citenamefont {Ferrie}, \citenamefont {Gross},\ and\ \citenamefont
  {Emerson}}]{Veitch2012-ld}%
  \BibitemOpen
  \bibfield  {author} {\bibinfo {author} {\bibfnamefont {V.}~\bibnamefont
  {Veitch}}, \bibinfo {author} {\bibfnamefont {C.}~\bibnamefont {Ferrie}},
  \bibinfo {author} {\bibfnamefont {D.}~\bibnamefont {Gross}},\ and\ \bibinfo
  {author} {\bibfnamefont {J.}~\bibnamefont {Emerson}},\ }\href@noop {}
  {\bibfield  {journal} {\bibinfo  {journal} {New J. Phys.}\ }\textbf {\bibinfo
  {volume} {14}},\ \bibinfo {pages} {113011} (\bibinfo {year}
  {2012})}\BibitemShut {NoStop}%
\bibitem [{\citenamefont {Walschaers}(2021)}]{Walschaers2021-ld}%
  \BibitemOpen
  \bibfield  {author} {\bibinfo {author} {\bibfnamefont {M.}~\bibnamefont
  {Walschaers}},\ }\href@noop {} {\bibfield  {journal} {\bibinfo  {journal}
  {PRX quantum}\ }\textbf {\bibinfo {volume} {2}} (\bibinfo {year}
  {2021})}\BibitemShut {NoStop}%
\bibitem [{\citenamefont {Kim}\ \emph {et~al.}(1989)\citenamefont {Kim},
  \citenamefont {{de Oliveira FA}},\ and\ \citenamefont {Knight}}]{Kim1989-ke}%
  \BibitemOpen
  \bibfield  {author} {\bibinfo {author} {\bibfnamefont {M.~S.}\ \bibnamefont
  {Kim}}, \bibinfo {author} {\bibnamefont {{de Oliveira FA}}},\ and\ \bibinfo
  {author} {\bibfnamefont {P.~L.}\ \bibnamefont {Knight}},\ }\href@noop {}
  {\bibfield  {journal} {\bibinfo  {journal} {Phys. Rev. A Gen. Phys.}\
  }\textbf {\bibinfo {volume} {40}},\ \bibinfo {pages} {2494} (\bibinfo {year}
  {1989})}\BibitemShut {NoStop}%
\bibitem [{\citenamefont {Moller}\ \emph {et~al.}(1996)\citenamefont {Moller},
  \citenamefont {Jorgensen},\ and\ \citenamefont {Dahl}}]{Moller1996-zo}%
  \BibitemOpen
  \bibfield  {author} {\bibinfo {author} {\bibfnamefont {K.~B.}\ \bibnamefont
  {Moller}}, \bibinfo {author} {\bibfnamefont {T.~G.}\ \bibnamefont
  {Jorgensen}},\ and\ \bibinfo {author} {\bibfnamefont {J.~P.}\ \bibnamefont
  {Dahl}},\ }\href@noop {} {\bibfield  {journal} {\bibinfo  {journal} {Phys.
  Rev. A}\ }\textbf {\bibinfo {volume} {54}},\ \bibinfo {pages} {5378}
  (\bibinfo {year} {1996})}\BibitemShut {NoStop}%
\bibitem [{\citenamefont {Nieto}(1997)}]{Nieto1997-rt}%
  \BibitemOpen
  \bibfield  {author} {\bibinfo {author} {\bibfnamefont {M.~M.}\ \bibnamefont
  {Nieto}},\ }\href@noop {} {\bibfield  {journal} {\bibinfo  {journal} {Phys.
  Lett. A}\ }\textbf {\bibinfo {volume} {229}},\ \bibinfo {pages} {135}
  (\bibinfo {year} {1997})}\BibitemShut {NoStop}%
\bibitem [{\citenamefont {Olivares}\ and\ \citenamefont
  {Paris}(2005)}]{Olivares2005-yw}%
  \BibitemOpen
  \bibfield  {author} {\bibinfo {author} {\bibfnamefont {S.}~\bibnamefont
  {Olivares}}\ and\ \bibinfo {author} {\bibfnamefont {M.~G.~A.}\ \bibnamefont
  {Paris}},\ }\href@noop {} {\bibfield  {journal} {\bibinfo  {journal} {J. Opt.
  B Quantum Semiclassical Opt.}\ }\textbf {\bibinfo {volume} {7}},\ \bibinfo
  {pages} {S616} (\bibinfo {year} {2005})}\BibitemShut {NoStop}%
\bibitem [{\citenamefont {Biswas}\ and\ \citenamefont
  {Agarwal}(2007)}]{Biswas2007-ni}%
  \BibitemOpen
  \bibfield  {author} {\bibinfo {author} {\bibfnamefont {A.}~\bibnamefont
  {Biswas}}\ and\ \bibinfo {author} {\bibfnamefont {G.~S.}\ \bibnamefont
  {Agarwal}},\ }\href@noop {} {\bibfield  {journal} {\bibinfo  {journal} {Phys.
  Rev. A}\ }\textbf {\bibinfo {volume} {75}} (\bibinfo {year}
  {2007})}\BibitemShut {NoStop}%
\bibitem [{\citenamefont {Ourjoumtsev}\ \emph {et~al.}(2006)\citenamefont
  {Ourjoumtsev}, \citenamefont {Tualle-Brouri}, \citenamefont {Laurat},\ and\
  \citenamefont {Grangier}}]{Ourjoumtsev2006-dt}%
  \BibitemOpen
  \bibfield  {author} {\bibinfo {author} {\bibfnamefont {A.}~\bibnamefont
  {Ourjoumtsev}}, \bibinfo {author} {\bibfnamefont {R.}~\bibnamefont
  {Tualle-Brouri}}, \bibinfo {author} {\bibfnamefont {J.}~\bibnamefont
  {Laurat}},\ and\ \bibinfo {author} {\bibfnamefont {P.}~\bibnamefont
  {Grangier}},\ }\href@noop {} {\bibfield  {journal} {\bibinfo  {journal}
  {Science}\ }\textbf {\bibinfo {volume} {312}},\ \bibinfo {pages} {83}
  (\bibinfo {year} {2006})}\BibitemShut {NoStop}%
\bibitem [{\citenamefont {Neergaard-Nielsen}\ \emph {et~al.}(2006)\citenamefont
  {Neergaard-Nielsen}, \citenamefont {Nielsen}, \citenamefont {Hettich},
  \citenamefont {M?lmer},\ and\ \citenamefont
  {Polzik}}]{Neergaard-Nielsen2006-mm}%
  \BibitemOpen
  \bibfield  {author} {\bibinfo {author} {\bibfnamefont {J.~S.}\ \bibnamefont
  {Neergaard-Nielsen}}, \bibinfo {author} {\bibfnamefont {B.~M.}\ \bibnamefont
  {Nielsen}}, \bibinfo {author} {\bibfnamefont {C.}~\bibnamefont {Hettich}},
  \bibinfo {author} {\bibfnamefont {K.}~\bibnamefont {M?lmer}},\ and\ \bibinfo
  {author} {\bibfnamefont {E.~S.}\ \bibnamefont {Polzik}},\ }\href@noop {}
  {\bibfield  {journal} {\bibinfo  {journal} {Phys. Rev. Lett.}\ }\textbf
  {\bibinfo {volume} {97}},\ \bibinfo {pages} {083604} (\bibinfo {year}
  {2006})}\BibitemShut {NoStop}%
\bibitem [{\citenamefont {Wakui}\ \emph {et~al.}(2007)\citenamefont {Wakui},
  \citenamefont {Takahashi}, \citenamefont {Furusawa},\ and\ \citenamefont
  {Sasaki}}]{Wakui2007-qj}%
  \BibitemOpen
  \bibfield  {author} {\bibinfo {author} {\bibfnamefont {K.}~\bibnamefont
  {Wakui}}, \bibinfo {author} {\bibfnamefont {H.}~\bibnamefont {Takahashi}},
  \bibinfo {author} {\bibfnamefont {A.}~\bibnamefont {Furusawa}},\ and\
  \bibinfo {author} {\bibfnamefont {M.}~\bibnamefont {Sasaki}},\ }\href@noop {}
  {\bibfield  {journal} {\bibinfo  {journal} {Opt. Express}\ }\textbf {\bibinfo
  {volume} {15}},\ \bibinfo {pages} {3568} (\bibinfo {year}
  {2007})}\BibitemShut {NoStop}%
\bibitem [{\citenamefont {Gerrits}\ \emph {et~al.}(2010)\citenamefont
  {Gerrits}, \citenamefont {Glancy}, \citenamefont {Clement}, \citenamefont
  {Calkins}, \citenamefont {Lita}, \citenamefont {Miller}, \citenamefont
  {Migdall}, \citenamefont {Nam}, \citenamefont {Mirin},\ and\ \citenamefont
  {Knill}}]{Gerrits2010-vr}%
  \BibitemOpen
  \bibfield  {author} {\bibinfo {author} {\bibfnamefont {T.}~\bibnamefont
  {Gerrits}}, \bibinfo {author} {\bibfnamefont {S.}~\bibnamefont {Glancy}},
  \bibinfo {author} {\bibfnamefont {T.~S.}\ \bibnamefont {Clement}}, \bibinfo
  {author} {\bibfnamefont {B.}~\bibnamefont {Calkins}}, \bibinfo {author}
  {\bibfnamefont {A.~E.}\ \bibnamefont {Lita}}, \bibinfo {author}
  {\bibfnamefont {A.~J.}\ \bibnamefont {Miller}}, \bibinfo {author}
  {\bibfnamefont {A.~L.}\ \bibnamefont {Migdall}}, \bibinfo {author}
  {\bibfnamefont {S.~W.}\ \bibnamefont {Nam}}, \bibinfo {author} {\bibfnamefont
  {R.~P.}\ \bibnamefont {Mirin}},\ and\ \bibinfo {author} {\bibfnamefont
  {E.}~\bibnamefont {Knill}},\ }\href@noop {} {\bibfield  {journal} {\bibinfo
  {journal} {arXiv [quant-ph]}\ } (\bibinfo {year} {2010})}\BibitemShut
  {NoStop}%
\bibitem [{\citenamefont {Takase}\ \emph {et~al.}(2021)\citenamefont {Takase},
  \citenamefont {Yoshikawa}, \citenamefont {Asavanant}, \citenamefont {Endo},\
  and\ \citenamefont {Furusawa}}]{Takase2021-co}%
  \BibitemOpen
  \bibfield  {author} {\bibinfo {author} {\bibfnamefont {K.}~\bibnamefont
  {Takase}}, \bibinfo {author} {\bibfnamefont {J.-I.}\ \bibnamefont
  {Yoshikawa}}, \bibinfo {author} {\bibfnamefont {W.}~\bibnamefont
  {Asavanant}}, \bibinfo {author} {\bibfnamefont {M.}~\bibnamefont {Endo}},\
  and\ \bibinfo {author} {\bibfnamefont {A.}~\bibnamefont {Furusawa}},\
  }\href@noop {} {\bibfield  {journal} {\bibinfo  {journal} {Phys. Rev. A}\
  }\textbf {\bibinfo {volume} {103}} (\bibinfo {year} {2021})}\BibitemShut
  {NoStop}%
\bibitem [{\citenamefont {Korolev}\ \emph {et~al.}(2023)\citenamefont
  {Korolev}, \citenamefont {Bashmakova}, \citenamefont {Tagantsev},\ and\
  \citenamefont {Golubeva}}]{Korolev2023-vs}%
  \BibitemOpen
  \bibfield  {author} {\bibinfo {author} {\bibfnamefont {S.~B.}\ \bibnamefont
  {Korolev}}, \bibinfo {author} {\bibfnamefont {E.~N.}\ \bibnamefont
  {Bashmakova}}, \bibinfo {author} {\bibfnamefont {A.~K.}\ \bibnamefont
  {Tagantsev}},\ and\ \bibinfo {author} {\bibfnamefont {T.~Y.}\ \bibnamefont
  {Golubeva}},\ }\href@noop {} {\bibfield  {journal} {\bibinfo  {journal}
  {arXiv [quant-ph]}\ } (\bibinfo {year} {2023})}\BibitemShut {NoStop}%
\bibitem [{\citenamefont {Bashmakova}\ \emph
  {et~al.}(2025{\natexlab{a}})\citenamefont {Bashmakova}, \citenamefont
  {Korolev}, \citenamefont {Zinatullin}, \citenamefont {Golubev},\ and\
  \citenamefont {Golubeva}}]{Bashmakova2025-tc}%
  \BibitemOpen
  \bibfield  {author} {\bibinfo {author} {\bibfnamefont {E.~N.}\ \bibnamefont
  {Bashmakova}}, \bibinfo {author} {\bibfnamefont {S.~B.}\ \bibnamefont
  {Korolev}}, \bibinfo {author} {\bibfnamefont {E.~R.}\ \bibnamefont
  {Zinatullin}}, \bibinfo {author} {\bibfnamefont {Y.~M.}\ \bibnamefont
  {Golubev}},\ and\ \bibinfo {author} {\bibfnamefont {T.~Y.}\ \bibnamefont
  {Golubeva}},\ }\href@noop {} {\bibfield  {journal} {\bibinfo  {journal} {J.
  Opt. Technol.}\ }\textbf {\bibinfo {volume} {92}},\ \bibinfo {pages} {195}
  (\bibinfo {year} {2025}{\natexlab{a}})}\BibitemShut {NoStop}%
\bibitem [{\citenamefont {Bashmakova}\ \emph
  {et~al.}(2025{\natexlab{b}})\citenamefont {Bashmakova}, \citenamefont
  {Korolev},\ and\ \citenamefont {Golubeva}}]{Bashmakova2025-lm}%
  \BibitemOpen
  \bibfield  {author} {\bibinfo {author} {\bibfnamefont {E.~N.}\ \bibnamefont
  {Bashmakova}}, \bibinfo {author} {\bibfnamefont {S.~B.}\ \bibnamefont
  {Korolev}},\ and\ \bibinfo {author} {\bibfnamefont {T.~Y.}\ \bibnamefont
  {Golubeva}},\ }\href@noop {} {\bibfield  {journal} {\bibinfo  {journal}
  {Phys. Rev. A}\ }\textbf {\bibinfo {volume} {112}} (\bibinfo {year}
  {2025}{\natexlab{b}})}\BibitemShut {NoStop}%
\bibitem [{\citenamefont {Hope}\ \emph {et~al.}(2025)\citenamefont {Hope},
  \citenamefont {Lidal},\ and\ \citenamefont {Massel}}]{Hope2025-vl}%
  \BibitemOpen
  \bibfield  {author} {\bibinfo {author} {\bibfnamefont {M.~K.}\ \bibnamefont
  {Hope}}, \bibinfo {author} {\bibfnamefont {J.}~\bibnamefont {Lidal}},\ and\
  \bibinfo {author} {\bibfnamefont {F.}~\bibnamefont {Massel}},\ }\href@noop {}
  {\bibfield  {journal} {\bibinfo  {journal} {arXiv [quant-ph]}\ } (\bibinfo
  {year} {2025})}\BibitemShut {NoStop}%
\bibitem [{\citenamefont {Lenzini}\ \emph {et~al.}(2018)\citenamefont
  {Lenzini}, \citenamefont {Janousek}, \citenamefont {Thearle}, \citenamefont
  {Villa}, \citenamefont {Haylock}, \citenamefont {Kasture}, \citenamefont
  {Cui}, \citenamefont {Phan}, \citenamefont {Dao}, \citenamefont {Yonezawa},
  \citenamefont {Lam}, \citenamefont {Huntington},\ and\ \citenamefont
  {Lobino}}]{Lenzini2018-su}%
  \BibitemOpen
  \bibfield  {author} {\bibinfo {author} {\bibfnamefont {F.}~\bibnamefont
  {Lenzini}}, \bibinfo {author} {\bibfnamefont {J.}~\bibnamefont {Janousek}},
  \bibinfo {author} {\bibfnamefont {O.}~\bibnamefont {Thearle}}, \bibinfo
  {author} {\bibfnamefont {M.}~\bibnamefont {Villa}}, \bibinfo {author}
  {\bibfnamefont {B.}~\bibnamefont {Haylock}}, \bibinfo {author} {\bibfnamefont
  {S.}~\bibnamefont {Kasture}}, \bibinfo {author} {\bibfnamefont
  {L.}~\bibnamefont {Cui}}, \bibinfo {author} {\bibfnamefont {H.-P.}\
  \bibnamefont {Phan}}, \bibinfo {author} {\bibfnamefont {D.~V.}\ \bibnamefont
  {Dao}}, \bibinfo {author} {\bibfnamefont {H.}~\bibnamefont {Yonezawa}},
  \bibinfo {author} {\bibfnamefont {P.~K.}\ \bibnamefont {Lam}}, \bibinfo
  {author} {\bibfnamefont {E.~H.}\ \bibnamefont {Huntington}},\ and\ \bibinfo
  {author} {\bibfnamefont {M.}~\bibnamefont {Lobino}},\ }\href@noop {}
  {\bibfield  {journal} {\bibinfo  {journal} {Sci. Adv.}\ }\textbf {\bibinfo
  {volume} {4}},\ \bibinfo {pages} {eaat9331} (\bibinfo {year}
  {2018})}\BibitemShut {NoStop}%
\bibitem [{\citenamefont {Eaton}\ \emph {et~al.}(2022)\citenamefont {Eaton},
  \citenamefont {Gonzalez-Arciniegas}, \citenamefont {Alexander}, \citenamefont
  {Menicucci},\ and\ \citenamefont {Pfister}}]{Eaton2022-ed}%
  \BibitemOpen
  \bibfield  {author} {\bibinfo {author} {\bibfnamefont {M.}~\bibnamefont
  {Eaton}}, \bibinfo {author} {\bibfnamefont {C.}~\bibnamefont
  {Gonzalez-Arciniegas}}, \bibinfo {author} {\bibfnamefont {R.~N.}\
  \bibnamefont {Alexander}}, \bibinfo {author} {\bibfnamefont {N.~C.}\
  \bibnamefont {Menicucci}},\ and\ \bibinfo {author} {\bibfnamefont
  {O.}~\bibnamefont {Pfister}},\ }\href@noop {} {\bibfield  {journal} {\bibinfo
   {journal} {Quantum}\ }\textbf {\bibinfo {volume} {6}},\ \bibinfo {pages}
  {769} (\bibinfo {year} {2022})}\BibitemShut {NoStop}%
\bibitem [{\citenamefont {Hope}\ \emph {et~al.}(2026)\citenamefont {Hope},
  \citenamefont {Lidal},\ and\ \citenamefont {Massel}}]{Hope2026-ry}%
  \BibitemOpen
  \bibfield  {author} {\bibinfo {author} {\bibfnamefont {M.~K.}\ \bibnamefont
  {Hope}}, \bibinfo {author} {\bibfnamefont {J.}~\bibnamefont {Lidal}},\ and\
  \bibinfo {author} {\bibfnamefont {F.}~\bibnamefont {Massel}},\ }\href@noop {}
  {\bibfield  {journal} {\bibinfo  {journal} {Phys. Rev. Res.}\ }\textbf
  {\bibinfo {volume} {8}} (\bibinfo {year} {2026})}\BibitemShut {NoStop}%
\bibitem [{\citenamefont {Gottesman}\ \emph {et~al.}(2001)\citenamefont
  {Gottesman}, \citenamefont {Kitaev},\ and\ \citenamefont
  {Preskill}}]{Gottesman2001-ke}%
  \BibitemOpen
  \bibfield  {author} {\bibinfo {author} {\bibfnamefont {D.}~\bibnamefont
  {Gottesman}}, \bibinfo {author} {\bibfnamefont {A.}~\bibnamefont {Kitaev}},\
  and\ \bibinfo {author} {\bibfnamefont {J.}~\bibnamefont {Preskill}},\
  }\href@noop {} {\bibfield  {journal} {\bibinfo  {journal} {Phys. Rev. A}\
  }\textbf {\bibinfo {volume} {64}},\ \bibinfo {pages} {012310} (\bibinfo
  {year} {2001})}\BibitemShut {NoStop}%
\bibitem [{\citenamefont {Weigand}\ and\ \citenamefont
  {Terhal}(2018)}]{Weigand2018-qe}%
  \BibitemOpen
  \bibfield  {author} {\bibinfo {author} {\bibfnamefont {D.~J.}\ \bibnamefont
  {Weigand}}\ and\ \bibinfo {author} {\bibfnamefont {B.~M.}\ \bibnamefont
  {Terhal}},\ }\href@noop {} {\bibfield  {journal} {\bibinfo  {journal} {Phys.
  Rev. A}\ }\textbf {\bibinfo {volume} {97}} (\bibinfo {year}
  {2018})}\BibitemShut {NoStop}%
\bibitem [{\citenamefont {Winnel}\ \emph {et~al.}(2024)\citenamefont {Winnel},
  \citenamefont {Guanzon}, \citenamefont {Singh},\ and\ \citenamefont
  {Ralph}}]{Winnel2024-om}%
  \BibitemOpen
  \bibfield  {author} {\bibinfo {author} {\bibfnamefont {M.~S.}\ \bibnamefont
  {Winnel}}, \bibinfo {author} {\bibfnamefont {J.~J.}\ \bibnamefont {Guanzon}},
  \bibinfo {author} {\bibfnamefont {D.}~\bibnamefont {Singh}},\ and\ \bibinfo
  {author} {\bibfnamefont {T.~C.}\ \bibnamefont {Ralph}},\ }\href@noop {}
  {\bibfield  {journal} {\bibinfo  {journal} {Phys. Rev. Lett.}\ }\textbf
  {\bibinfo {volume} {132}},\ \bibinfo {pages} {230602} (\bibinfo {year}
  {2024})}\BibitemShut {NoStop}%
\end{thebibliography}%
	
\end{document}